\documentclass[pra,twocolumn]{revtex4}
\usepackage{graphicx}
\usepackage{dcolumn}
\usepackage{bm}
\usepackage{amsmath}
\usepackage{amsfonts}
\usepackage[usenames]{color}
\usepackage{epsfig}
\usepackage{epstopdf} 
\DeclareGraphicsExtensions{.pdf,.eps,.png,.jpg,.mps}

\begin{document}
\title{Carnot's theorem for nonthermal stationary reservoirs}
 
\author{Simone \surname{De Liberato}}
\author{Masahito Ueda}
\affiliation{Department of Physics, University of Tokyo, Hongo, Bunkyo-ku, Tokyo 113-0033, Japan}

\begin{abstract}
Carnot's theorem poses a fundamental limit on the maximum efficiency achievable from an engine that works between two reservoirs at thermal equilibrium. 
We extend this result to the case of arbitrary nonthermal stationary reservoirs, even with quantum coherence.
In order to do this we prove that  a single nonthermal reservoir is formally equivalent to multiple equilibrium ones.
Finally, we discuss  the possibility of realizing an engine that works at unit efficiency by exploiting quantum coherence present in the reservoir.
\end{abstract}
\maketitle

 \section{Introduction}
For almost two centuries, Carnot's theorem has constituted one of the cornerstones of thermodynamics, setting a fundamental bound on the efficiency of any heat-to-work conversion process. It states that:
{\it All reversible engines working between two reservoirs at temperatures $T_C$ and $T_H$ have the same efficiency $\eta_C=1-\frac{T_C}{T_H}$. No engine working between two reservoirs at thermal equilibrium can have an efficiency greater than that.}
Here the efficiency is defined as the ratio between $W$, the work performed by the engine, and $Q_H$, the heat extracted from the hotter reservoir.

In this paper we generalize Carnot's theorem to a more general setting in which the reservoirs are not in thermal equilibrium (and thus $T_C$ and $T_H$ cannot be defined). In the field of quantum thermodynamics, examples of such nonthermal reservoirs can be found, for example, in the study of  engines with strongly coupled  \cite{Nori2} or quantum coherent \cite{Scully} reservoirs. While in the following we will concentrate on such kind of {\it microscopic} examples,  it is worthwhile to notice that, in general, most of the engines present in our everyday world actually extract energy from nonequilibrium environments (e.g., all living being extracting energy from ATP molecules).

This paper is structured as follows: in Section \ref{genthe} we will start by introducing our approach to heat engines that we will then use in Section \ref{carnotst}  to rederive the standard Carnot's theorem. In Section \ref{eqthe} 
we will prove a general equivalence theorem, stating that a nonthermal reservoir is formally equivalent to a collection of equilibrium ones at different temperatures. This result will be used in Section \ref{noneqcarnot} to prove a generalized version of the Carnot's theorem, valid for general  nonthermal reservoirs. In Section \ref{scully} such theorem will be tested against some previously known result.
Finally, using the developed theory, in Section \ref{unit}, we discuss the possibility of realizing an engine exploiting  quantum coherence to work at unit efficiency. Conclusion and perspectives are drawn in Section \ref{conc}.

\begin{figure}[t!]
\begin{center}
\includegraphics[width=8cm]{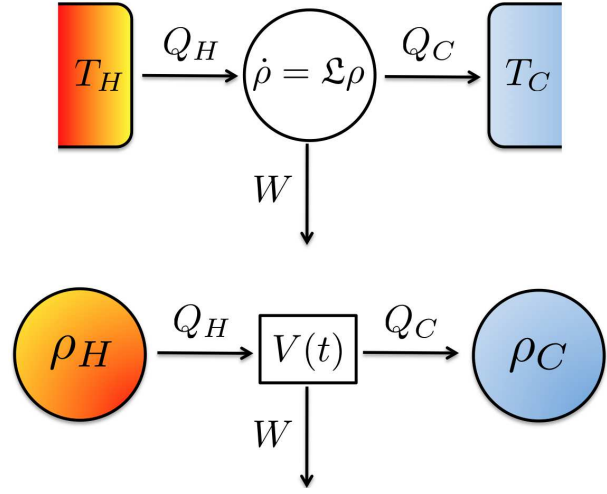}
\caption{ \label{Sketch} Top panel: the standard approach in the study of heat engines. The engine's degrees of freedom evolve under the influence of fixed reservoirs (e.g., through some form of Liouvillian operator $\mathfrak{L}$). Bottom panel: the approach proposed in this paper. The reservoirs represent the dynamical degrees of freedom of the theory, and they evolve under the effect of  coupling $V(t)$ mediated by the engine.}
\end{center}
\end{figure}

\section{General theory}
\label{genthe}
In order to develop our program we are immediately confronted with the problem of defining what we mean with {\it reservoir}.
In order to keep our theory as general as possible, we will consider the largest possible definition of reservoir, that is, we will consider a reservoir to be a completely general physical system, with the only constraint  that its size and the strength of the interaction with the external world (the engine in our case) are such that its state  is not affected sensibly by the time evolution. 

The generality of the above definition implies that we disregard any {\it a priori} difference between a heat reservoir and a work reservoir. The difference will only be {\it a posteriori}. If our theory implies that it is possible to extract work by coupling the engine with only one reservoir, we will classify it as a work reservoir. A typical example would be an inverted system, such that a higher energy level has a population bigger than a lower lying one. It is well known that energy can be extracted from such systems without any need of using a second reservoir (e.g., a laser extracting energy from an inverted medium). 

While the formalism we will develop could be used to describe such systems, it is not meaningful to define an efficiency in the usual sense for them (by energy conservation $W=Q_H$ and thus $\eta=1$). We will thus disregard such systems because Carnot's theorem does not apply to them.
When, in the following, we will talk about {\it heat} engines, we will imply engines working with reservoirs defined in such a sense.

To formulate our theory in a model-independent manner, we adopt a  slightly unusual approach in the study of heat engines. 
Heat engines are normally studied taking into consideration some of their degrees of freedom, evolving under the influence of two (or eventually more) external reservoirs. Given that an engine is, by definition, cyclical, after a cycle the engine is back to its initial state while the reservoirs have slightly evolved.
We can thus model the action of the engine as an operator coupling the reservoirs  and allowing energy flows between them (see Fig. \ref{Sketch} for a schematic representation of the two approaches).
This point of view will allow us to consider arbitrary reservoirs, be they  at thermal equilibrium or not, and to optimize the efficiency of the engine over the space of all the possible interaction operators (i.e., all the possible engines).

In our theory, the reservoirs are thus dynamical objects and we describe them in terms of their density operators $\rho_H$ and $\rho_C$ and Hamiltonians $H_H$ and $H_C$, whose eigenvalues we call $E_H$ and $E_C$. Given that the reservoirs can {\it a priori} be  nonthermal, the subscripts $H$ (hot) and $C$ (cold) have no direct implication of their temperatures, but rather are used to differentiate the energy source (hot  reservoir) and drain (cold reservoir). 
In order to be able to derive our main results, we will restrict our attention to the case in which the initial states of the decoupled reservoirs are time independent, that is
\begin{eqnarray}
\label{comm}
[H_C,\rho_C]=[H_H,\rho_H]&=&0.
\end{eqnarray} 

The engine's effective role is to couple the two reservoirs. It can thus be described completely by a Hermitian time-dependent coupling operator $\lambda V(t)$, where $\lambda \in \mathbb{R}$ and $V(t)$ is an operator over the tensor product of the Hilbert spaces of the two reservoirs (without the time dependence the engine would conserve the total energy of the reservoirs and thus extract no work).  

In the definition of {\it reservoir} we chose, it is explicitly required that its state does not change in any significant way during the interaction with the engine. For this reason, to recover usual thermodynamic results from our approach, we have to consider the limit of  vanishing interaction $\lambda\rightarrow 0$, thus developing the theory to the first nonvanishing order in $\lambda$.  This limit is well defined because the efficiency, which is  given by the ratio between work and heat fluxes, will not depend on $\lambda$. 

In the interaction picture, the Liouville equation for the system, up to the second order, takes the form
\begin{eqnarray}
\label{master}
\dot{\rho}(t)&=&i\lambda[\rho(0),\tilde{V}(t)]-\lambda^2 \int_{0}^{t}  [[\rho(0),\tilde{V}(\tau)],\tilde{V}(t)] d\tau, \quad  \quad
\end{eqnarray}
where $\tilde{V}(t)=e^{it(H_H+H_C)}V(t)e^{-it(H_H+H_C)}$ is the perturbation in the  interaction picture
and $\rho(0)=\rho_H \otimes \rho_C$ is the initial density matrix.
In order to calculate the heat flow from reservoir $j=\{C,H\}$, we can use the quantum version of the first law of thermodynamics \cite{Nori}.
The total internal energy of reservoir $j$ is given by
\begin{eqnarray}
\label{U0}
U_j= \text{Tr} (\rho(t) H_j),
\end{eqnarray}
and thus its time variation is
\begin{eqnarray}
\label{TU}
\dot{U}_j= \text{Tr} (\rho(t) \dot{H}_j)+\text{Tr} (\dot{\rho}(t) H_j).
\end{eqnarray}
The two terms in the right-hand side of Eq. \ref{TU} can be identified respectively with the exchanged work and heat, 
making Eq. \ref{TU}  a quantum version of the first law of thermodynamics. The interested reader is invited to read Ref. \cite{Nori} and references
therein for a discussion of the relevance of such identification.
In our case the Hamiltonian of the reservoirs is time independent and thus they exchange no mechanical work.
The exchanged heat, which in the present case is equal to the total energy variation, can thus be calculated as
\begin{eqnarray}
\label{flow}
 \dot{Q}_j(t)=-\text{Tr} (\dot{\rho}(t) H_j),
 \end{eqnarray}
where  we have chosen the convention that $Q_j$ is positive if heat is extracted {\it from} the reservoir.
Inserting Eq. \ref{master} into Eq. \ref{flow} we have
\begin{eqnarray}
\label{flow2}
 \dot{Q}_j(t)&=&-i\lambda\text{Tr} ([\rho(0),\tilde{V}(t)]H_j)\nonumber \\&&+\lambda^2 \int_{0}^{t}  \text{Tr}([[\rho(0),\tilde{V}(\tau)],\tilde{V}(t)]H_j) d\tau ,
 \end{eqnarray}
 where the first term in the right-hand side vanishes due to Eq. \ref{comm}.
Formally integrating Eq. \ref{flow2} up to final time $t_f$, chosen to be a multiple of the engine period,  we obtain the total amount of heat exchanged with each reservoir
\begin{eqnarray}
\label{heat}
 Q_j&=&\frac{\lambda^2}{2} \text{Tr}([[\rho(0),M],M]H_j ), 
\end{eqnarray}
where  
\begin{eqnarray}
M=\int_0^{t_f} \tilde{V}(t) dt,
\end{eqnarray}
and we have exploited the fact that, using the Jacobi identity and the fact that the initial states of the decoupled reservoirs are time-independent (Eq. \ref{comm}), we have, for $\forall t_1,t_2$, 
\begin{eqnarray}
\text{Tr}([[\rho(0),\tilde{V}(t_1)],\tilde{V}(t_2)]H_j)=\text{Tr}([[\rho(0),\tilde{V}(t_2)],\tilde{V}(t_1)]H_j).\,
\end{eqnarray}
 The net balance of energy between the two reservoirs gives the total work extracted by the engine 
\begin{eqnarray}
\label{work}
W&=&Q_H+Q_C.
\end{eqnarray}
Introducing indexes $(p,q)$ over the  energy eigenstates of the cold reservoir and $(m,n)$ over the eigenstates of the hot one, and noticing that, because of Eq. \ref{comm}, $\rho_C$ and $\rho_H$ can be made diagonal in such a basis,  we can rewrite Eq. \ref {heat} elementwise as 
\begin{eqnarray}
\label{heat_elem}
Q_C&=& \lambda^2\sum_{m,n,p,q} \lvert M_{mp}^{nq} \lvert^2 (\rho^m_H\rho^p_C-\rho^n_H\rho^q_C)E_C^p,\nonumber\\
Q_H&=& \lambda^2\sum_{m,n,p,q} \lvert M_{mp}^{nq} \lvert^2 (\rho^m_H\rho^p_C-\rho^n_H\rho^q_C)E_H^m.
\end{eqnarray}
Exploiting the hermiticity of $M$ ($ \lvert M_{mp}^{nq} \lvert= \lvert M_{nq}^{mp} \lvert$), we can rewrite Eq. \ref{heat_elem}  summing only over states such that $E_H^m > E_H^n$, thus obtaining
\begin{eqnarray}
\label{heat_elem_half}
Q_C&=& \lambda^2\sum_{\substack{m,n,p,q \\ E_H^m>E_H^n}} \lvert M_{mp}^{nq} \lvert^2 (\rho^m_H\rho^p_C-\rho^n_H\rho^q_C)(E_C^p-E_C^q),\nonumber\\
Q_H&=& \lambda^2\sum_{\substack{m,n,p,q \\ E_H^m>E_H^n}} \lvert M_{mp}^{nq} \lvert^2 (\rho^m_H\rho^p_C-\rho^n_H\rho^q_C)(E_H^m-E_H^n).
\nonumber\\
\end{eqnarray}
Equation \ref{heat_elem_half} gives a self-contained, model independent description of energy exchanges in a general heat engine. The explicit dependance over the Hamiltonian spectrum makes this formalism particularly adapted to the study 
of the engine efficiency at the ultimate quantum limit \cite{Popescu, Popescu2, Linden}.
 
\section{Standard Carnot's theorem}
\label{carnotst}
It is interesting to notice that the standard Carnot's theorem can be easily derived from Eq. \ref{heat_elem_half}, by choosing
properly normalized thermal distributions for the reservoirs
\begin{eqnarray}
\label{res} 
\rho_H^m&=&e^{- E_H^m/T_H}/Z_H,\quad
\rho_C^p=e^{- E_C^p/T_C}/Z_C,
\end{eqnarray}
with $T_H\geq T_C$.
In order to have the engine extract heat from the hot reservoir ($Q_H\geq 0$), from Eq. \ref{heat_elem_half}  we need to have  the condition
\begin{eqnarray}
\label{cond}
\rho^m_H\rho^p_C-\rho^n_H\rho^q_C&\geq&0,
\end{eqnarray}
verified at least for some values of $(m,n,p,q)$ (please refer to Appendix \ref{appendix} for a detailed justification of this important point).
Using the reservoirs in Eq. \ref{res}, Eq. \ref{cond} becomes 
\begin{eqnarray}
\label{thercond} 
\frac{E_C^q-E_C^p}{E_H^m-E_H^n}\geq \frac{T_C}{T_H}.
\end{eqnarray}
Writing down the engine efficiency using Eqs. \ref{work} and  \ref{heat_elem_half}, we have
\begin{eqnarray}
\label{extended}
\eta&=&\frac{W}{Q_H}=1+\frac{Q_C}{Q_H}\\&=&1-\frac{\sum_{\substack{m,n,p,q \\ E_m>E_n}} \lvert M_{mp}^{nq} \lvert^2 (\rho^m_H\rho^p_C-\rho^n_H\rho^q_C)(E_C^q-E_C^p)}{ \sum_{\substack{m,n,p,q \\ E_m>E_n}} \lvert M_{mp}^{nq} \lvert^2 (\rho^m_H\rho^p_C-\rho^n_H\rho^q_C)(E_H^m-E_H^n)}\nonumber.
\end{eqnarray}
It is easy to verify that any value of $(m,n,p,q)$ that does not satisfy Eq. \ref{cond}  lowers the overall efficiency $\eta$. We can thus rewrite Eq. \ref{extended} as
\begin{eqnarray}
\label{extended2}
\eta&\leq&1-\frac{\sum^{'}_{\substack{m,n,p,q \\ E_m>E_n}} \lvert M_{mp}^{nq} \lvert^2 (\rho^m_H\rho^p_C-\rho^n_H\rho^q_C)(E_C^q-E_C^p)}{ \sum^{'}_{\substack{m,n,p,q \\ E_m>E_n}} \lvert M_{mp}^{nq} \lvert^2 (\rho^m_H\rho^p_C-\rho^n_H\rho^q_C)(E_H^m-E_H^n)},\nonumber \\
\end{eqnarray}
where the prime over the sum symbol means we sum only over indexes such that Eq. \ref{cond} is satisfied.

By construction, each term in the sum in the numerator of Eq. \ref{extended2} obeys
the inequality in Eq. \ref{thercond}. We can thus fix a lower bound on each of the terms  and obtain Carnot's result
\begin{eqnarray}
\label{premaineff}
\eta&\leq& 1-\frac{T_C}{T_H}.
\end{eqnarray}

If all the transitions take place between almost equilibrium states, the left-hand side of Eq. \ref{cond} tends toward zero and Eqs. \ref{thercond} and \ref{premaineff} become equalities. This is independent from the chosen engine interaction $M_{mp}^{nq}$, because each term $\lvert M_{mp}^{nq} \lvert^2 (\rho^m_H\rho^p_C-\rho^n_H\rho^q_C)$ in Eq. \ref{heat_elem_half}, being present both in $Q_C$ and $Q_H$, simplifies in Eq. \ref{premaineff}.
That is, consistently with the usual formulation of Carnot's theorem, we find that any engine working between states almost at equilibrium, attains Carnot efficiency 
\begin{eqnarray}
\eta_C=1-\frac{T_C}{T_H},
\end{eqnarray}
(a similar approach to the Carnot efficiency has recently been proposed in \cite{Popescu}).
The fact that we can prove Carnot's theorem from our formalism is not surprising, because systems at thermal equilibrium (like the reservoirs in  Eq. \ref{res}) are known to obey it. This is a good consistency check for our approach.

\section{Reservoir equivalence theorem}
\label{eqthe}
If the reservoirs' distributions differ from thermal equilibrium ones, we cannot in general define a temperature for them and thus  Eq. \ref{premaineff} does not apply.  In this section we will prove a reservoir equivalence theorem that we will use in Section \ref{noneqcarnot} to establish a generalized form of  Carnot's theorem, valid for arbitrary nonthermal reservoirs.  

We start by rewriting the sum in Eq. \ref{heat_elem_half} as a sum over all the possible transitions between pairs of states in each reservoir
\begin{eqnarray}
\label{heat_elem_half_doub}
Q_C&=& \lambda^2\sum_{\substack{m\rightarrow n\\ E_m>E_n}}\sum_{p\rightarrow  q} \lvert M_{mp}^{nq} \lvert^2 (\rho^m_H\rho^p_C-\rho^n_H\rho^q_C)(E_C^p-E_C^q),\nonumber\\
Q_H&=& \lambda^2\sum_{\substack{m\rightarrow n\\ E_m>E_n}}\sum_{p\rightarrow  q} \lvert M_{mp}^{nq} \lvert^2 (\rho^m_H\rho^p_C-\rho^n_H\rho^q_C)(E_H^m-E_H^n).
\nonumber\\
\end{eqnarray}
From Eq. \ref{heat_elem_half_doub}, we see that the heat flow between the two reservoirs is composed of the sum over all the possible pairwise interactions,  coupling a transition in the cold reservoir (from $p$ to $q$) and a transition in the hot one (from $m$ to $n$).
This means that, modulo a renormalization of the density operator (that only amounts to a redefinition of the engine interaction $M_{mp}^{nq}$), the heat flow between two reservoirs with multiple levels (and thus multiple transitions) is formally equivalent to the flow between multiple reservoirs, each one with only two levels (and thus only one transition).
A single engine working between the cold and hot reservoirs is thus equivalent  to a set of different engines, each one working between two reservoirs composed respectively of the two-level systems made of the levels $(p,q)$ and $(m,n)$.

\begin{figure}[t!]
\begin{center}
\includegraphics[width=9cm]{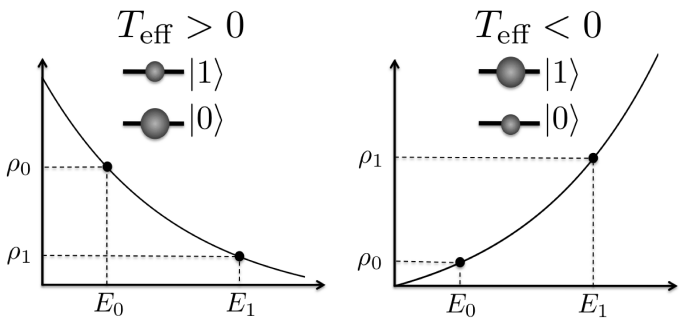}
\caption{ \label{exp} An arbitrary two-level system, with level populations  $\rho_0$ and $\rho_1$ and level energies $E_0$ and $E_1$, can always be considered in thermal equilibrium for a certain effective temperature $T_{\text{eff}}$. This temperature is
positive if the lower lying level population is higher (left panel) or negative otherwise (right panel).}
\end{center}
\end{figure}

This point is important for us because a two-level system  with arbitrary level populations  $\rho_0$ and $\rho_1$ and level energies $E_0$ and $E_1$, can always be considered in thermal equilibrium for a certain effective temperature $T_{\text{eff}}$. This is due to the elementary calculus result that, given two arbitrary points in the Cartesian plane, there is always an exponential function connecting them (see Fig. \ref{exp} for a schematic illustration). The effective temperature  $T_{\text{eff}}$, that can be positive or negative, will thus be given by the equation
\begin{eqnarray}
\label{twolev}
\frac{\rho_1}{\rho_0}&=&e^{-(E_1-E_0)/T_\text{eff}}.
\end{eqnarray}

From the two remarks above we obtain one of our main result:
{\it An arbitrary time-independent reservoir is formally equivalent to a collection of equilibrium sub-reservoirs composed of two-level systems, each one characterized by its effective equilibrium temperature given by Eq. \ref{twolev}.}
 
Given that a reservoir is usually made up of many identical subsystems (e.g., the molecules of a gas), each transition will be highly degenerate. Through Eq. \ref{twolev} this will give rise to many identical two-level systems, leading to sub-reservoirs roughly of the same size (in terms of the number of subsystems and thus of the number of available transitions) of the original ones.
It is important to note that the equivalence is purely {\it formal}; these sub-reservoirs are only mathematical objects, useful abstractions. We are not suggesting that the reservoir is phisically divided into multiple pieces. 
It is also important to remember that the hypothesis that the initial states  are stationary states of the decoupled reservoirs (Eq. \ref{comm}) is required for the theorem to hold.

\section{Generalzed Carnot's theorem}
\label{noneqcarnot}
\begin{figure}[t!]
\begin{center}
\includegraphics[width=8cm]{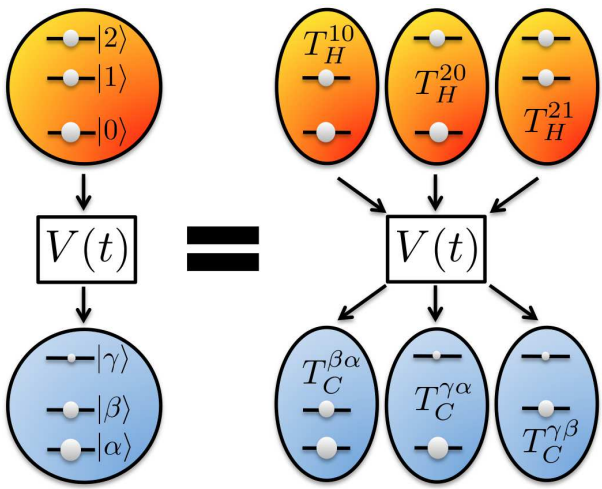}
\caption{ \label{Equiv} An engine working between two nonthermal reservoirs is formally
equivalent to an engine working between two sets of  thermal sub-reservoirs. }
\end{center}
\end{figure}

From the theorem proven in the previous section, the generalization of Carnot's theorem we are looking for follows quite naturally.
Being a nonthermal reservoir equivalent to a collection of equilibrium sub-reservoirs,
an engine working between two of them is in fact formally equivalent to one operating between two sets of equilibrium sub-reservoirs, each one with its own effective temperature. The engine couples pairs of sub-reservoirs, one from the cold side and the other from the hot one, extracting work from them (see Fig. \ref{Equiv} for a schematic illustration in the case of two reservoirs composed of three-level systems). 
We have now a system that can be described using only thermal reservoirs and we can thus easily find an upper bound to its efficiency using the standard Carnot theorem and the tools we developed in Section \ref{carnotst}. 
Using Eq. \ref{twolev}  we can define the effective, transition-dependent temperatures for each pair of levels as
\begin{eqnarray}
\label{temps}
T_C^{qp}&=&(E_C^q-E_C^p)/\log \frac{\rho_C^p}{\rho_C^q},\\
T_H^{mn}&=&(E_H^m-E_H^n)/\log \frac{\rho_H^n}{\rho_H^m}\nonumber,
\end{eqnarray}
and, following the standard Carnot theorem, or better its trivial expansion to the case of multiple reservoirs, 
we obtain the following upper bound on the efficiency of energy extraction from the  nonthermal reservoirs 
\begin{eqnarray}
\label{effneq}
\eta\leq1-\frac{\min(T_C^{qp})}{\max(T_H^{mn})},
\end{eqnarray}
where the minimum and the maximum are taken respectively over all the pairs of levels in the cold $(p,q)$ and hot $(m,n)$ reservoirs.

In particular the engine described by an operator $M_{mp}^{nq}$, whose only nonzero elements have indexes $(\tilde{m},\tilde{n})$ and  $(\tilde{p},\tilde{q})$ that satisfy
\begin{eqnarray}
\label{tildas}
T_H^{\tilde{m}\tilde{n}}&=&\max(T_H^{mn}),\\
T_C^{\tilde{q}\tilde{p}}&=&\min(T_C^{qp})\nonumber,
\end{eqnarray}
couples only pairs of transitions corresponding to the lowest and highest temperature, that is, it works only between the coldest and the hottest reservoir. It will thus obey Eq. \ref{effneq} and it can saturate the inequality if all the transitions take place between almost equilibrium states, as explained in Section \ref{carnotst}.

In order to write Eq. \ref{effneq} we made two supplementary hypotesis:  all the effective temperatures are positive and all the effective temperatures of the hot reservoir are hotter than those of the cold one. If the first hypotesis is violated, it could be possible to construct an engine extracting work from a single reservoir; if the second is violated, we could extract work from bidirectional heat flows.
In both cases, while our formalism is completely apt to study them, the usual definition of efficiency is not well suited (see the discussion about work reservoirs in Section \ref{genthe})
and thus it is meaningless to apply the Carnot's theorem, and we will thus ignore these possibilities. 
We have thus proved that the determination of the upper bound of the efficiency can be reduced to the calculation of the extrema of the effective temperatures in the two reservoirs. 

While we will consider in the following only examples concerning systems with discrete spectra, the same procedure can be generalized to the generic continuum case. The effective temperatures in each reservoir will then form two dimensional surfaces and their extrema can be located by usual analytic or numerical methods.

\section{Application to a known case}
\label{scully}
In the final part of this paper,  we will apply the theory just developed to study the efficiency that can be obtained from reservoirs presenting some amount of quantum coherence.
This case was treated in a paper by Scully and coworkers  \cite{Scully}. In this paper they  showed how, given a reservoir consisting of a thermal gas of three-level atoms with a certain amount of quantum coherence between the quasi-degenerate two lower levels, it is possible to build an engine with an efficiency greater than the one given by Carnot's theorem.
In order to do that, they devised a Photo-Carnot engine whose working fluid is composed of photons, that uses the thermal three-level quantum coherent atom gas as a hot reservoir and a generic reservoir at the same temperature, but without coherence, as a cold one. 
We will show how our theory allows us to find the same results in a complete model-independent way (that is without any need of devising an actual engine). 

Following Ref. \cite{Scully} we will define a thermal, quantum coherent system as a system whose density matrix has diagonal elements given by thermal populations and some nonzero off-diagonal terms.
The coherent gas is thus described by the density matrix
\begin{eqnarray}
\label{rhophi}
\rho_{\phi}= \left( \begin{array}{ccc}
P_a & 0 & 0 \\ 0 & P_b & \rho_{bc}e^{i\phi} \\ 0 & \rho_{bc}e^{-i\phi} & P_c
\end{array} \right),
\end{eqnarray}
where the diagonal elements are the thermal populations of the three states. In the following we will consider the degenerate case $P_b=P_c$, in order to satisfy Eq. \ref{comm} and be able to apply our equivalence theorem,
and we will call $\Omega$ the energy gap between the higher level and the lower two.
In the limit of high temperature and small coherence, Scully and coworkers find an efficiency for the Photo-Carnot engine depending on the phase between the two coherent levels, given by
\begin{eqnarray}
\label{etascu}
\eta_{\phi}=-\frac{P_a \rho_{bc} \cos(\phi)}{P_b(P_b-P_a)},
\end{eqnarray}
where, given the two reservoirs at the same temperature, we would expect a zero efficiency in the absence of coherence.
To apply our theory we diagonalize the density matrix in Eq. \ref{rhophi},
obtaining the eigenvalues $[P_a,  P_b-\rho_{bc}, P_b+ \rho_{bc}]$. 
The thermal, coherent gas, is thus equivalent to a fully incoherent, but nonthermal gas. 
Applying Eq. \ref{temps} we find the following three effective temperatures for the hot reservoir
\begin{eqnarray}
\label{Tp}
T_H^{ab}&=&\Omega / \log\left(\frac{P_b-\rho_{bc}}{P_a}\right),\nonumber \\
T_H^{ac}&=&\Omega / \log\left(\frac{P_b+\rho_{bc}}{P_a}\right),\\
T_H^{bc}&=&0\nonumber,
\end{eqnarray}
while for the incoherent, cold reservoir, we have a single, equilibrium temperature
\begin{eqnarray}
\label{Tnp}
T_C&=&\Omega / \log\left(\frac{P_b}{P_a}\right).
\end{eqnarray}
Substituting Eqs. \ref{Tp} and \ref{Tnp} into Eq. \ref{effneq}, we obtain the maximal efficiency given by
\begin{eqnarray}
\label{maxmio}
\eta&\leq&1-\frac{\log(\frac{P_b-\rho_{bc}}{P_a})}{\log(\frac{P_b}{P_a})},
\end{eqnarray}
which, in the high temperature ($P_b\simeq P_a$) and small coherence ($\rho_{bc}\ll1$) regime, reduces to
\begin{eqnarray}
\label{etascumio}
\eta \leq \frac{P_a \rho_{bc}}{P_b(P_b-P_a)},
\end{eqnarray}
that is the maximum of Eq. \ref{etascu} (actually following the  calculations in Ref. \cite{Scully}, but without making any simplifying approximation, we would find the optimal efficiency exactly as in Eq. \ref{maxmio}).
We have thus proved that our theory can correctly predict, in a model-independent way, the maximal efficiency 
of the Photo-Carnot engine. Moreover we have shown that the efficiency found in Ref. \cite{Scully} is indeed optimal for the chosen cold and hot reservoirs.

\section{Unit efficiency engine}
\label{unit}
While proving the results of the previous section, we stumbled upon a rather unexpected result. One of the temperature in Eq. \ref{Tp}, corresponding to the transition between the two coherent degenerate levels, has an effective temperature equal to $0$. This seems to imply that, switching the two reservoir, that is using the coherent reservoir as the cold one, it should be possible to conceive an engine working at unit efficiency (the $T_C$ in Eq. \ref{effneq} is equal to $0$ and thus $\eta=1$).

Initially puzzling, 
this turns out to be a generic feature of reservoirs with degenerate, coherent levels. Two coherent degenerate levels are generally described by a density matrix of the form 
\begin{eqnarray}
\label{rhoc}
\rho_{c}=\frac{1}{2} \left( \begin{array}{cc}
1 & \sigma  \\ \sigma & 1 
\end{array} \right),
\end{eqnarray}
that, after diagonalization, yields  an effective zero temperature, as can be seen from 
Eq. \ref{temps}, because the energies are equal while the populations are different. 

The physical origin of such seemingly unphysical behavior is easy to understand.
The entropy of $\rho_c$ is always lower than that of the fully incoherent density matrix 
\begin{eqnarray}
\label{rhoi}
\rho_{i}= \frac{1}{2}\left( \begin{array}{cc}
1 & 0  \\ 0 & 1 
\end{array} \right).
\end{eqnarray}
Since all the states in such a degenerate subspace have the same energy, the reservoir can act as a perfect  
entropy drain, absorbing entropy but not energy from the engine as it evolves from $\rho_c$ to $\rho_i$.
We thus predict the possibility to realize an engine with unit efficiency, extracting work from a single reservoir and dissipating entropy by destroying coherence in a second, coherent reservoir.
The maximal efficiency of such engine would be independent of the strength of the coherence $\sigma$, but
the work extractable from it would depend on the total amount of coherence that is burned by the engine.
A simple application of the second law of thermodynamics gives the following upper bound 
\begin{eqnarray}
\label{slow}
W\leq T_H M \Delta S,
\end{eqnarray}
where $M$ is the total number of pairs of levels whose coherence is utilized to extract work $W$ and $\Delta S$ is the entropy difference between $\rho_i$ and $\rho_c$. 
 While such predictions of high efficiencies might seem to violate the usual Carnot bound, this is not the case. In fact, as we stated above, we are considering the efficiency of 
 work extraction from reservoirs already in a nonthermal state. In order to compare such results with the ones obtained for thermal reservoirs  we should also consider the processes needed to bring the reservoirs out of thermal equilibrium in the first place. Illuminating discussions in the case  of the Photo-Carnot engine can be found in  \cite{Zubairy,Sun}.

The above construction is rather formal, but there are  simple and well-known systems that implement the mechanism described.
Without entering into detailed calculations, that are beyond the scope of this paper, we can notice that a laser without inversion \cite{ScullyBook} driven by a thermal field can indeed behave as a unit efficiency engine. If the active medium is composed of coherent, degenerate $\Lambda$ atoms, whose upper level is at an energy high enough to be able to neglect its thermal occupation, such a system effectively extracts energy (in the form of laser radiation) from the thermal reservoir (the thermal field) with unit efficiency, while destroying the coherence of the $\Lambda$ atoms.

It is also interesting to notice that destroying coherence between degenerate levels is not the only way to eliminate entropy without loosing energy. A number of investigations have indeed shown, exploiting the links between thermodynamics and information theory  \cite{I1,I2,I3,I4,I5,I6,I7}, that if information is somehow extracted from the system, it is possible to obtain  an engine working at unit efficiency.

\section{Conclusions and perspectives}
\label{conc}

In the present paper we have introduced a new approach to  study the efficiency of thermal engines that allows us to treat general nonthermal and quantum coherent reservoirs.
We have proved that our approach gives the same result as the canonical one when applied to thermal situations  (Section \ref{carnotst})  and yields the right result 
when applied to the only case of a nonthermal reservoir thoroughly studied in literature (Section \ref{noneqcarnot}).
While the formalism we developed differs in many respects from what usually done in standard thermodynamics, the fact that it is able to give the right results in all known situations
 vindicates the correctness of our approach.

As applications we reproduced the results originally derived in Ref. \cite{Scully} and then showed the possibility of realizing an engine with unit efficiency that exploits quantum coherence

The aforementioned results were derived using rather formal methods. It will be an interesting challenge  to identify and study physical systems in which such results can be tested and applied.

\section{Acknowledgments} 
We acknowledge useful discussions with I. Carusotto, S. Endo, N. Lambert and K. Maruyama.
This work was supported by KAKENHI (22340114 and 22103005), Global COE Program‚ the Physical Sciences Frontier and the Photon Frontier Network Program, MEXT, Japan. SDL acknowledges FY2009 JSPS Postdoctoral Fellowship for Foreign Researchers.

\appendix
\section{Proof of Equation \ref{cond}}
\label{appendix}

In Section \ref{carnotst} we claimed that, in order to have an engine that extracts heat from the hot reservoir, we need to have 
\begin{eqnarray}
\label{acond}
\rho^m_H\rho^p_C-\rho^n_H\rho^q_C&>&0,
\end{eqnarray}
at least for some value of the indexes $(m,n,p,q)$. Here we will give a detailed justification of this point. 

Following what is done in Section \ref{carnotst} we can write the heat fluxes from the two reservoirs as
\begin{eqnarray}
\label{aheat_elem_half}
Q_C&=& \lambda^2\sum_{\substack{m,n,p,q \\ E_H^m>E_H^n}} \lvert M_{mp}^{nq} \lvert^2 (\rho^m_H\rho^p_C-\rho^n_H\rho^q_C)(E_C^p-E_C^q),\nonumber\\
Q_H&=& \lambda^2\sum_{\substack{m,n,p,q \\ E_H^m>E_H^n}} \lvert M_{mp}^{nq} \lvert^2 (\rho^m_H\rho^p_C-\rho^n_H\rho^q_C)(E_H^m-E_H^n),
\nonumber\\
\end{eqnarray}
and we assume the reservoirs to be thermal 
\begin{eqnarray}
\label{ares} 
\rho_H^m&=&e^{- E_H^m/T_H}/Z_H,\quad
\rho_C^p=e^{- E_C^p/T_C}/Z_C,
\end{eqnarray}
with $T_H > T_C$.
The explicit form we derived for $Q_C$ and $Q_H$ in Eq. \ref{aheat_elem_half}
allows us to write the heat fluxes between the reservoirs and the engine as a sum over different channels, indexed by the 4-tuple $(m,n,p,q)$.
If we examine the contribution of each channel to $Q_C$ and $Q_H$, we can realize that,
in order for the channel to extract some heat from the reservoirs, we need two conditions to be fulfilled:
\begin{enumerate}
\item at least one of the two contributions has to be positive;
\item if only one contribution is positive, its norm has to be bigger that the norm of the other.
\end{enumerate}
If these two conditions are not fulfilled, the channel is effectively dissipating work into the reservoirs, and thus lowering the overall efficiency.
In order to have an engine extracting some work we thus need these two conditions to be fulfilled at least for some channel.

In the rest of this Appendix we are going to prove that any channel fulfilling the two conditions has $Q_H>0$ and $Q_C<0$, and thus it verifies Eq. \ref{acond}.
We will prove this by showing that any other possibility leads to contradiction.

\vspace{0.6cm}

\subsection{$Q_H>0$ and $Q_C>0$}

Having both contributions from  Eq. \ref{aheat_elem_half} positive would imply
\begin{eqnarray}
\label{aconde}
\rho^m_H\rho^p_C-\rho^n_H\rho^q_C&>&0,
\end{eqnarray}
and 
\begin{eqnarray}
\label{acondbis}
E_C^p-E_C^q&>&0.
\end{eqnarray}
Yet, using the reservoirs in Eq. \ref{ares}, Eq. \ref{aconde} implies 
\begin{eqnarray}
\label{athercond} 
\frac{E_C^q-E_C^p}{E_H^m-E_H^n}> \frac{T_C}{T_H},
\end{eqnarray}
that is never verified, as the left-hand side is negative.
We have thus proved that it is not possible for an engine to extract heat from both reservoirs.

\subsection{$Q_H<0$ and $Q_C>0$}

This would imply, from Eq.  \ref{aheat_elem_half},
\begin{eqnarray}
\label{acondn}
\rho^m_H\rho^p_C-\rho^n_H\rho^q_C&<&0.
\end{eqnarray}
Condition (2) thus imposes the following inequality
\begin{eqnarray}
\label{acondnorm}
E_C^q-E_C^p&>&E_H^m-E_H^n.
\end{eqnarray}
From Eqs. \ref{acondn} and \ref{acondnorm}, using the reservoirs in Eq. \ref{ares}, we obtain the following relation
\begin{eqnarray}
\label{thercondimp} 
1<\frac{E_C^q-E_C^p}{E_H^m-E_H^n}< \frac{T_C}{T_H},
\end{eqnarray}
which is also never verified.
We have thus proved that it is not possible for an engine to work extracting heat from the cold reservoir.

\subsection{$Q_H<0$ and $Q_C<0$}

This case would trivially violate condition (1), as both contributions are negative.

\end{document}